\documentclass[prb,superscriptaddress,longbibliography,twocolumn]{revtex4-1}
\usepackage{geometry}
\usepackage{amsmath}
\usepackage{amssymb}
\usepackage{graphicx}
\usepackage{hyperref}
\usepackage{braket}
\usepackage{bm}
\usepackage[usenames,dvipsnames]{color}

\graphicspath{./}

\begin{document}

\title{Topological Mechanics from Supersymmetry}

\author{Jan Attig}
\affiliation{Institute for Theoretical Physics, University of Cologne, 50937 Cologne, Germany}
\author{Krishanu Roychowdhury}
\affiliation{Laboratory of Atomic And Solid State Physics, Cornell University, Ithaca, NY 14853, USA}
\affiliation{Department of Physics, Stockholm University, SE-106 91 Stockholm, Sweden}
\author{Michael J. Lawler}
\affiliation{Laboratory of Atomic And Solid State Physics, Cornell University, Ithaca, NY 14853, USA}
\affiliation{Department of Physics, Binghamton University, Binghamton, NY, 13902, USA}
\author{Simon Trebst}
\affiliation{Institute for Theoretical Physics, University of Cologne, 50937 Cologne, Germany}

\begin{abstract}
In topological mechanics, the identification of a mechanical system's rigidity matrix with an electronic 
tight-binding model allows to infer topological properties of the mechanical system, such as the occurrence
of `floppy' boundary modes, from the associated electronic band structure. Here we introduce an 
approach to systematically construct topological mechanical systems 
by an exact supersymmetry (SUSY)  
that relates the bosonic (mechanical) and fermionic (e.g.~electronic) degrees of freedom. 
As examples we discuss mechanical analogues of the Kitaev honeycomb model and of a second-order topological 
insulator with floppy corner modes. Our SUSY construction naturally defines hitherto unexplored topological 
invariants for bosonic (mechanical) systems, such as bosonic Wilson loop operators that are formulated in terms of 
a SUSY-related fermionic Berry curvature.

\end{abstract}
\date{\today}

\maketitle

%%%%%%%%%%%%%%%%%%%%%%%%%%%%%%%%%%%%%%%%%%%%%%%%%%%%%%
% INTRODUCTION
%%%%%%%%%%%%%%%%%%%%%%%%%%%%%%%%%%%%%%%%%%%%%%%%%%%%%%

In quantum mechanics, supersymmetry (SUSY) explicitly relates bosonic and fermionic degrees of freedom
-- a fundamental concept that has first been introduced \cite{Golfand1971,Ramond1971,Neveu1971} 
in high-energy physics and widely been adopted in the formulation of extensions of the standard model 
\cite{StandardModel}.
For non-relativistic settings, the concept of SUSY has been exploited extensively 
in the study of random phenomena and quantum chaos in mesoscopic systems \cite{Efetov1999},
where a supersymmetric combination of commuting and anticommuting variables allows for the otherwise intractable calculation of disorder averages.
The discovery of topological (classical) mechanics has produced another intriguing setting 
where two fundamentally distinct degrees of freedom are found to be closely related -- the {\em mechanical}
modes of a classical system are cast in analogy to the wavefunction of an {\em electronic} system 
\cite{Huber2016}.
At its heart, this analogy identifies the Newtonian equation governing the classical system, 
$
    \ddot{x} = - {\bf D} x,
$
with a Schr\"odinger equation
\begin{equation}
	i \frac{\partial}{\partial t}
	\begin{pmatrix} \sqrt{\mathbf{D}}^T x \\ i \dot{x} \end{pmatrix}
 	=
 	\begin{pmatrix}0 & \sqrt{\mathbf{D}}^T \\ \sqrt{\mathbf{D}} & 0 \end{pmatrix}
 	\begin{pmatrix} \sqrt{\mathbf{D}}^T x \\ i \dot{x} \end{pmatrix} \,,
	\label{eq:Schroedinger}
\end{equation}
where the dynamical matrix ${\bf D}$ of the original classical system enters the Hamiltonian matrix
of the quantum system.
Exploiting such matrix analogies has produced some far-reaching insight, most prominently in the realization
that zero-energy `floppy' boundary modes in isostatic lattices \cite{Maxwell1864} can be identified 
with the inherently protected boundary modes of topological insulators \cite{KaneLubensky2013}.
It has previously been noted \cite{KaneLubensky2013,Huber2016} that the Hamiltonian matrix in \eqref{eq:Schroedinger}
corresponds to a certain symmetry class (BDI) \cite{Altland1997} and therefore generically has a supersymmetric form \cite{Efetov1999}.
In going one step further, the concept of supersymmetry can indeed be used to explicitly connect the 
degrees of freedom underlying the aforementioned matrix analogy, which are intrinsically bosonic for the
mechanical system and fermionic for the eletronic system \cite{Lawler2016}.
More precisely, considering that the mechanical system is described in terms of real-space coordinates and
momenta, $(q,p)$, which are both {\em real} bosonic variables, their natural SUSY partners are not complex fermions but
{\em real} fermions, i.e. Majorana fermions.

It is the purpose of this manuscript to formulate an explicit SUSY construction which, 
for a broad family of topological Majorana fermion systems, will allow to build their topological mechanical counterparts
in terms of (simple) balls and springs models. 
As examples of this construction, 
we discuss mechanical incarnations of the Z$_2$ spin liquid phase in the Kitaev honeycomb model
and of a second-order topological insulator. Furthermore, this SUSY construction allows to 
identify topological properties of the bosonic (mechanical) system by explicitly associating 
it with a fermionic Berry curvature. 
We showcase such a calculation by evaluating bulk topological invariants, i.e.~bosonic Wilson loops,
for the mechanical equivalent of a second-order topological insulator. 
Our results demonstrate that one can construct hitherto unexplored topological mechanical systems by using the concept of supersymmetry to directly 
relate many of the recent advances from the field of topological quantum matter to mechanical analogues.

%%%%%%%%%%%%%%%%%%%%%%%%%%%%%%%%%%%%%%%%%%%%%%%%%%%%%%%%%%%%%%%%%%%%
% SUSY Construction
%%%%%%%%%%%%%%%%%%%%%%%%%%%%%%%%%%%%%%%%%%%%%%%%%%%%%%%%%%%%%%%%%%%%

\noindent {\em SUSY Construction.--}
To set the scene for our SUSY construction of topological mechanical systems, let us briefly 
recapitulate the concept of supersymmetry. The central object here is a non-hermitian SUSY charge operator
\begin{equation}
	\mathcal{Q} = c_i^\dagger{\bf R}_{ij} b_j^{\phantom\dagger}
	\label{eq:ComplexSUSYCharge}
\end{equation}
that connects the annihilation operator of a (complex) boson $b$ with the creation operator
of a (complex) fermion $c$ via an arbitrary matrix ${\bf R}$.
The indices $i$ and $j$ run over the total number of 
fermionic and bosonic degrees of freedom, respectively.
From this charge operator one can immediately construct a supersymmetric Hamiltonian
\begin{equation}
	\mathcal{H}_{\text{SUSY}} 
	=
	\{\mathcal{Q} , \mathcal{Q}^\dagger \}
	=
	 c^\dagger {\bf R R^\dagger} c +
	 b^\dagger {\bf R^\dagger R} b  
\end{equation}
that decomposes into decoupled bosonic and fermionic parts. 
By construction, these two partner Hamiltonians are not only both quadratic and isospectral to one another, 
but their eigenstates are explicitly related by $\mathcal{Q}$, allowing a one-to-one
identification of bosonic and fermionic states \footnote{More precisely, such a one-to-one
identification is possible only for non-zero energy eigenstates.}.

For the case of topological mechanics, with phase-space coordinates $(q,p)$ for bosonic degrees of freedom, we have to further specialize this SUSY connection to
the case of {\em real} bosons and real fermions (and later take the classical limit). 
More explicitly, we are now led to consider a hermitian SUSY charge
\begin{equation}
	\mathcal{Q}
	= 
	\gamma_i^B \mathbf{1}_{ij} \hat{p}_j + \gamma_i^A \mathbf{A}_{ij} \hat{q}_j 
	\label{eq:RealSUSYCharge}
\end{equation}
that connects the two bosonic degrees of freedom, $(\hat{p},\hat{q})$ 
(where the hats indicate that these are still quantum mechanical operators fulfilling the usual commutator relations $[\hat{q}_i,\hat{p}_j] = i \delta_{i,j}$), 
with a matching number of two species of Majorana fermions $\gamma^A$ and $\gamma^B$.
Note that in comparison to the complex boson/fermion case of Eq.~\eqref{eq:ComplexSUSYCharge}
we have restricted the matrix {\bf R} to a block-diagonal form
$
	{\bf R}
	=
	\left(\begin{smallmatrix}
		\mathbf{1} & \mathbf{0} \\
		\mathbf{0} & \mathbf{A}
	\end{smallmatrix}\right)
$
for reasons that 
are apparent when looking at the SUSY Hamiltonian $H_{\rm SUSY} = \{\mathcal{Q} , \mathcal{Q}^\dagger \}$ 
that again decomposes into bosonic and fermionic partner Hamiltonians,  taking the form
\begin{eqnarray} 
	\mathcal{H}_\text{fermion}  & =  & i \gamma_{j}^{B}\;\mathbf{A}_{jk}^T \;\gamma_{k}^{A} - i \gamma_{j}^{A}\;\mathbf{A}_{jk} \;\gamma_{k}^{B} \,,
	\label{eq:MFHamiltonian}
	\\
	\mathcal{H}_\text{boson} & = & \hat{p}_i \hat{p}_i + \hat{q}_i (\mathbf{A}^T \mathbf{A})_{ij} \hat{q}_j \,.
	\label{eq:BosonHamiltonian} 
\end{eqnarray}
Written in this way, the Majorana Hamiltonian \eqref{eq:MFHamiltonian} describes the hopping of Majorana fermions 
between two types of sites $A$ and $B$, as it is realized, for instance, for nearest-neighbor hopping on bipartite lattices. 
In such a two-sublattice realization, the bosonic operators reside only on one of the sublattices (namely $B$).

Taking the classical limit of the bosonic Hamiltonian \eqref{eq:BosonHamiltonian}, one can further read off that in this form ${\bf R}$ in fact corresponds to the rigidity matrix of the mechanical system,
with its upper left block giving rise to the mass matrix (set to unity here) and the lower right block giving rise to the dynamical matrix via ${\bf D} = \mathbf{A}^T \mathbf{A}$.
To summarize these steps, we have accomplished that by restricting the matrix ${\bf R}$ in the SUSY charge for the real boson/fermion case \eqref{eq:RealSUSYCharge} it can not only be interpreted as the rigidity matrix of the classical system, but it also connects the mechanical system
to a particularly accessible form of Majorana hopping problems. 
In more practical terms, our particular choice of ${\bf R}$ allows us to directly connect a number of well-known
Majorana systems [of form \eqref{eq:MFHamiltonian}]  to mechanical analogues as given by Eq.~\eqref{eq:BosonHamiltonian}.

%%%%%%%%%%%%%%%%%%%%%%%%%%%%%%%%%%%%%%%%%%%%%%%%%%%%%%%%%%%%%%%%%%%%
% Classical balls and springs models
%%%%%%%%%%%%%%%%%%%%%%%%%%%%%%%%%%%%%%%%%%%%%%%%%%%%%%%%%%%%%%%%%%%%

\noindent {\em Classical balls and springs models.--}
The final step in our SUSY construction is to translate the classical limit of the real boson model \eqref{eq:BosonHamiltonian} 
into a classical balls and springs model whose Hamiltonian can be written as
\begin{eqnarray}
	\mathcal{H} \; &=& \; \sum_i \frac{p_i^2}{2m} + \sum_{ij} \frac{k_{ij}}{2} (q_i - q_j)^2 + \sum_i \frac{\kappa_i}{2} q_i^2  \nonumber \\
    \; &\sim & \;\sum_i p_i^2 + \sum_{ij}q_i \mathbf{D}_{ij} q_j \,,
	\label{eq:boson-springs-general}
\end{eqnarray}
where the spring constants $k_{ij}$ and $\kappa_i$ can be extracted from the dynamical (spring) matrix $\mathbf{D} = \mathbf{A}^T\mathbf{A}$
as
\begin{eqnarray}
	k_{ij} & = & -2 \sum_{a \in A} \mathbf{A}^T_{i a} \mathbf{A}_{a j} \,, \label{eq:OffDiagonalSpring} \\
	\kappa_i & = & 2 \sum_{a \in A} {\bf A}_{i a}^2  \; - \; \sum_{b \in B} k_{i b} \,. \label{eq:DiagonalSpring}
\end{eqnarray}
The intersite spring constants $k_{ij}$ are the off-diagonal elements of ${\bf D}$, that by virtue of our SUSY construction 
arise from next-nearest neighbor Majorana hopping (within the boson sublattice $B$)
\footnote{For the common case of a uniform sign structure of the Majorana hopping terms, the intersite springs have {\em negative} spring constants, 
while the onsite spring constants are positive.}.
The onsite spring constants $\kappa_i$ are the diagonal elements of ${\bf A}^T {\bf A}$ arising from Majoranas hopping back-and-forth, modified by a contribution coming from the intersite springs.
 
%%%%%%%%%%%%%%%%%%%%%%%%%%%%%%%%%%%%%%%%%%%%%%%%%%%%%%%%%%%%%%%%%%%%
% Mechanical Kitaev model
%%%%%%%%%%%%%%%%%%%%%%%%%%%%%%%%%%%%%%%%%%%%%%%%%%%%%%%%%%%%%%%%%%%%

\begin{figure}[t]
	\centering
	\includegraphics[width=\columnwidth]{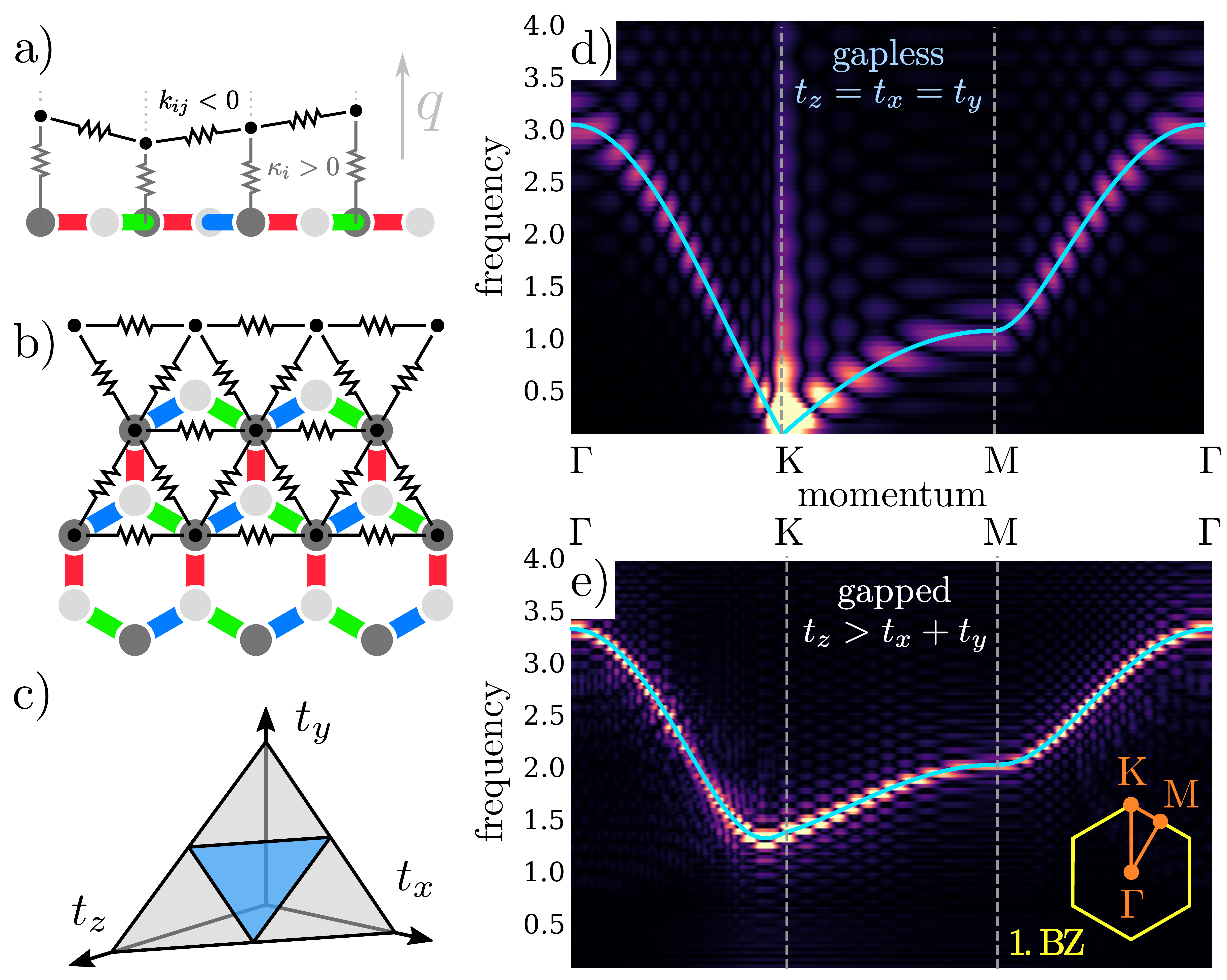}
	\caption{{\bf Mechanical Kitaev model.} 
			a) and b) Realization in the form of a classical balls \& springs model (see also main text).
			c) The phase diagram of the classical model exhibits a gapless region (blue) and three gapped phases (grey).
			The excitation spectra of the classical model extracted from numerical simulations for the d) gapless and e) gapped phases.
}
		\label{fig:kitaev}
\end{figure}

\noindent {\em Mechanical Kitaev model.--}
We now proceed to discuss a classical analogue of the celebrated Kitaev model \cite{Kitaev2006}, 
a spin model with characteristic bond-directional exchanges on the honeycomb lattice.
The analytical solution of this model\cite{Kitaev2006} is achieved by recasting it in terms of non-interacting Majorana fermions hopping 
on the same honeycomb lattice (in the background of a classical (static) $Z_2$ gauge field) -- precisely the type of Majorana Hamiltonian \eqref{eq:MFHamiltonian} 
that is amenable to our SUSY construction. 
Going through the steps outlined above, we end up with a classical balls and springs model on the triangular lattice
\footnote{Our model can also be considered to be a particularly simple balls and springs incarnation of ``mechanical graphene'', which has been first put forward in Ref.~\onlinecite{socolar2017mechanical}.}
(i.e. one of the two sublattices of the 
honeycomb lattice) as illustrated in Fig.~\ref{fig:kitaev}~a) and b). Each mass, located at a site of the triangular lattice, is restricted to a movement along an axis
perpendicular to the lattice plane, and is connected via two types of springs to both the plane and its neighboring masses.

To illustrate the physics of this mechanical Kitaev model, we have  
integrated the Hamiltonian equations of motion arising from \eqref{eq:boson-springs-general} for a system of $40\times40$ masses. By applying a periodic drive of a given frequency $\omega$ to a single mass located at the center of the system, we are able to probe individual eigenmodes of the mechanical model. 
To do so, we take real-space snapshots of the balls and springs configuration (for drives of different frequencies $\omega$), which we subsequently Fourier-transform. 
This allows us to recover the full energy dispersion of the classical system as shown in Figs.~\ref{fig:kitaev}~d) and e) for two sets of coupling parameters. In Fig.~\ref{fig:kitaev}~d) we probe the isotropically coupled
Kitaev model and recover the well known Dirac cone spectrum of the quantum system. Obtaining such a {\em linear} low-energy spectrum in a spring system (which on the level of individual springs always exhibits quadratic energy dispersions) is striking evidence of the many-body physics at play. 
In Fig.~\ref{fig:kitaev}~e) we show an energy spectrum for a situation where one of the three coupling parameters dominates and the spectrum exhibits a well-defined low-energy gap, i.e. the mechanical system remains rigid for low frequency drives
up to a threshold given by the gap.  
While this is imposed from the physics of the quantum system, it is again an unusual situation for a classical system, which typically defy a small-frequency rigidity (in particular on the level of individual springs)
\footnote{Note that for the system at hand, there are no gapless acoustic phonons, i.e. Goldstone modes arising from the spontaneous breaking of translational symmetry, since our system explicitly breaks this symmetry.}.  

The propagating phonon modes constitute the classical analogs of the Majorana fermions in the Kitaev model, with their energy spectra being in one-to-one correspondence. Note that also the underlying $Z_2$ gauge structure of the Kitaev spin liquid is fully present in the mechanical model.
A pair of gauge excitations -- visons in the language of  $Z_2$ spin liquids -- can be excited by flipping the sign of 
an intersite spring constant \footnote{Note that any change to the intersite spring constant also requires a compensating change to the onsite spring constants.},
in direct analogy to flipping the hopping on a bond in the quantum model. 
In total, our SUSY construction allows to build  a full mechanical analog of the $Z_2$ quantum spin liquid of the Kitaev model, 
complete with  classical analogs of both the fractional quasiparticles (Majorana fermions) and the underlying $Z_2$ lattice gauge structure.

%%%%%%%%%%%%%%%%%%%%%%%%%%%%%%%%%%%%%%%%%%%%%%%%%%%%%%%%%%%%%%%%%%%%
% Mechanical SOTI
%%%%%%%%%%%%%%%%%%%%%%%%%%%%%%%%%%%%%%%%%%%%%%%%%%%%%%%%%%%%%%%%%%%%
 
\begin{figure}[b]
	\centering
	\includegraphics[width=\columnwidth]{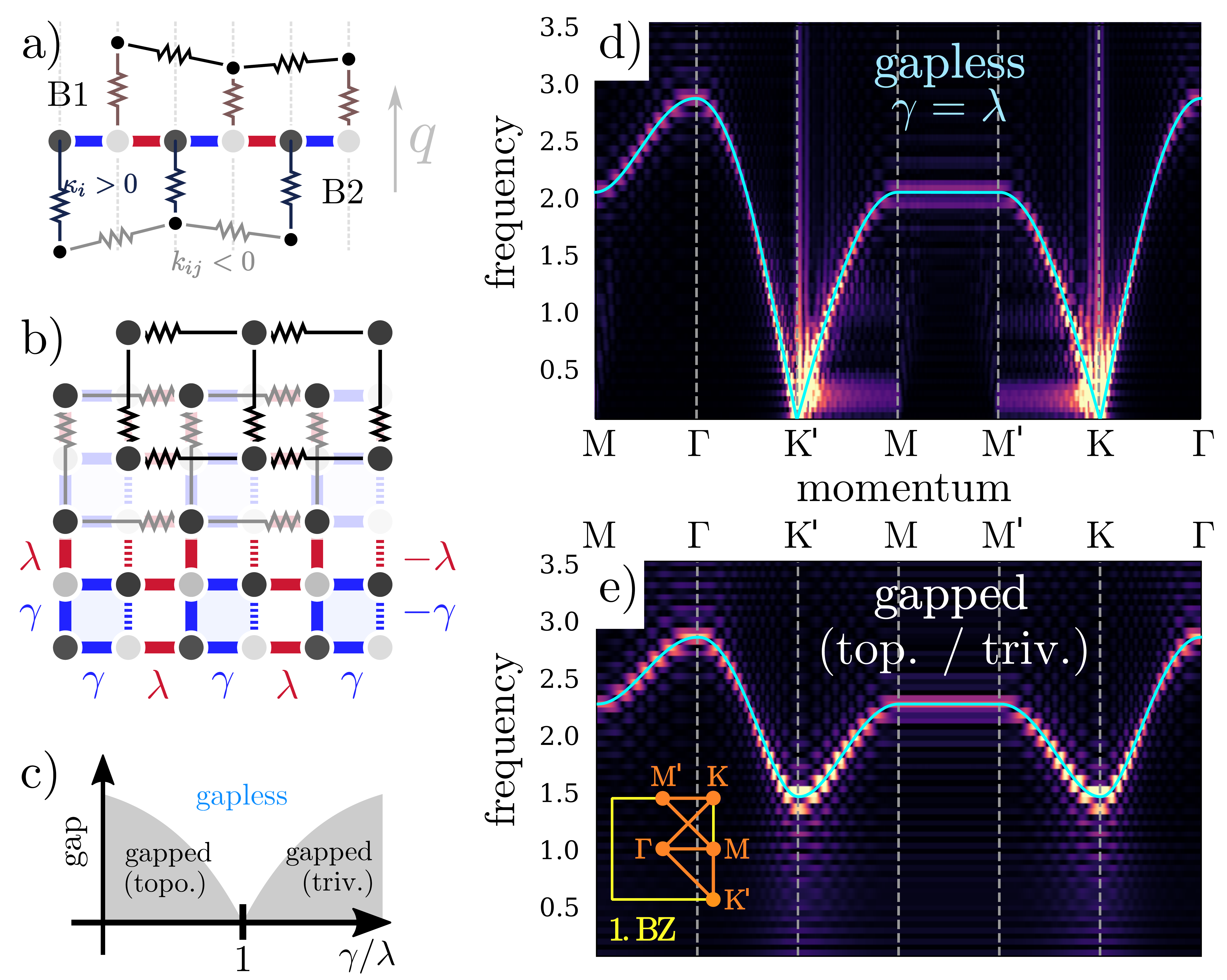}
	\caption{{\bf Balls \& springs model of a second-order topological insulator.}  Mechanical realization shown in a) side view and b) top view.
		       	As discussed in the main text the system decouples into two independent systems, denoted here by B1 and B2.
			c) Schematic phase diagram for a staggering of the coupling constants around the isotropic coupling point.
			 The excitation spectra of the classical model extracted from numerical simulations for the d) gapless and e) gapped (both topological and trivial) phase.
			 	\label{fig:SOTI}
			 }
\end{figure}

\noindent {\em Mechanical second-order TI.--}
As a second example, we apply our SUSY construction to derive a classical balls and springs model of the ``octupolar insulator'' introduced in Refs.~\onlinecite{hughes-bernevig_quadrupole_insulators,hughes-bernevig_multipole_insulators} as a principal example of a second-order topological insulator (SOTI) with topologically protected, gapless corner modes
\footnote{Conceptually, SOTIs are part of a family of higher-order topological insulators with protected hinge or corner modes \cite{neupert_HOTI}.
Experimental realizations of SOTIs have been discussed earlier this year in the context of cleverly engineered 
phononic insulators \cite{huber_HOTI_phonons}, 
microwave systems \cite{bahl_HOTI_microwave}, and
electrical circuits \cite{thomale_HOTI_electrical}, 
along with the observation that bismuth is in fact a SOTI \cite{neupert_HOTI_bismuth}. 
On a more theoretical basis, SOTIs have also been discussed in the context of frustrated quantum magnetism \cite{vatsal_HOTI_magnet}.
}. 
The original formulation \cite{hughes-bernevig_quadrupole_insulators} of the SOTI is based on a square lattice tight-binding model whose hopping strengths are staggered for the elementary square plaquettes of the lattice (which each encompasses a $\pi$-flux). While the original model is not sensitive to whether the underlying degrees of freedom are complex or real fermions, we again take the real-fermion formulation as principal input for our SUSY construction.  Going through the two steps of first constructing the SUSY-related real boson model \eqref{eq:BosonHamiltonian} and then taking its classical limit \eqref{eq:boson-springs-general}, we arrive at the balls and springs model illustrated in Figs.~\ref{fig:SOTI}~a) and b). 
The mechanical system is composed of two square lattices of coupled balls and springs (denoted B1 and B2 in the figure). 
The two lattices turn out to be decoupled, since any interlattice coupling always arises from {\em two} exchange paths in the quantum model that exactly cancel
(as the coupling along the two paths always involves opposite signs as mandated by the plaquette $\pi$-flux in the quantum model).
Concentrating on just one of the two mechanical lattices, this leaves us with a system of balls and springs where the masses are again restricted to move
along an out-of-plane axis only, with all intersite spring couplings taking the same value (independent of the staggering in the original quantum model).
The onsite spring couplings, on the other hand, will be crucial in realizing the topological corner modes of interest here. By construction [Eq.~\eqref{eq:DiagonalSpring}], the onsite couplings are sensitive to the number of neighbors. 
Considering a system with open boundary conditions, this gives rise to a spatial variation of these couplings along the boundary of the mechanical system.

Probing the mechanical system, we again demonstrate a one-to-one correspondence of its bulk energy spectra to the fermionic dispersions as illustrated for different staggerings in Figs.~\ref{fig:SOTI}~d) and e). The occurrence of gapless corner modes in the balls and springs system, predicted by analogy to the quantum model, can be probed by applying a small force to one of the corner masses. While in the coupling regime corresponding to the trivial phase of 
the quantum model there is a restoring force in the mechanical system, this is not the case when entering the coupling regime corresponding to the topological phase in the quantum model (i.e. for $\gamma/\lambda<1$). In this regime, the corner mass can be moved arbitrarily far from its original position when exercising an infinitesimal force -- this is the gapless corner mode in the mechanical system. 
That this mechanical corner mode is indeed the signature of a {\em bulk} topological phase in the classical system, can be made apparent by considering bulk topological invariants of the mechanical system as discussed in the following.

%%%%%%%%%%%%%%%%%%%%%%%%%%%%%%%%%%%%%%%%%%%%%%%%%%%%%%%%%%%%%%%%%%%%
% Topological invariants
%%%%%%%%%%%%%%%%%%%%%%%%%%%%%%%%%%%%%%%%%%%%%%%%%%%%%%%%%%%%%%%%%%%%

\noindent{\em Topological invariants.--}
A unique aspect of our SUSY construction is that it also defines a way to explore topological properties of mechanical/bosonic 
systems by connecting the symplectic bosonic eigenfunctions with a {\it fermionic} Berry phase of its SUSY partner.
The SUSY formalism thereby provides a platform to define {\em bosonic} invariants that characterize the topology of an arbitrary mechanical model, specified solely by its rigidity matrix, that go beyond the classification based on conventional bosonic Berry phases
(which are not capable of revealing the topological order). 

Consider a SUSY pair of fermion and boson models.
For the fermionic case, the Berry connection is given by
\begin{equation}\label{eq:fermionberry}
  {\mathcal A} = \langle u_m({\bf k})|i\nabla_k| u_n({\bf k})\rangle \,,
\end{equation}
where $|u_m({\bf k})\rangle$ are the eigenvectors of the fermionic Hamiltonian. These eigenstates map to the bosonic eigenstates $|v_m({\bf k})\rangle$ via the rigidity matrix \eqref{eq:ComplexSUSYCharge}
\begin{equation}
|u_m({\bf k})\rangle =\frac{{\bf R}({\bf k})}{\sqrt{|\omega_m({\bf k})|}} |v_m({\bf k})\rangle \equiv \tilde {\bf R}({\bf k})|v_m({\bf k})\rangle \,,
\end{equation}
with the prefactor ensuring the symplectic normalization $\langle v_m({\bf k})|\sigma_2|v_n({\bf k})\rangle = (\sigma_3)_{m,n}$ of the bosonic eigenproblem (see appendix~\ref{app:topologyofcornermodes}). Inserting into \eqref{eq:fermionberry} then leaves us with a definition of the fermionic Berry curvature in terms of the bosonic states
\begin{eqnarray}
	{\mathcal A}_{\rm SUSY} &=& \langle v_m({\bf k})|i \tilde{\bf R}^{\dagger}\nabla_k \left(\tilde{\bf R}|v_n({\bf k})\rangle\right) \nonumber \\
	&=& \langle v_m({\bf k})|i\sigma_2\left( \nabla_k +\sigma_2\tilde{\bf R}^{\dagger}\nabla_k \tilde{\bf R}\right)|v_n({\bf k})\rangle \,.
	\label{eq:SUSYBerryCurvature}
\end{eqnarray}
Note that the SUSY construction adds a covariant derivative to the conventional definition of a Berry curvature for the bosonic eigenstates. Importantly, bosonic eigenstates that are trivial with regard to the conventional definition can be revealed as topological states using this augmented definition.

\begin{figure}
\centering
\includegraphics[width=0.96\columnwidth]{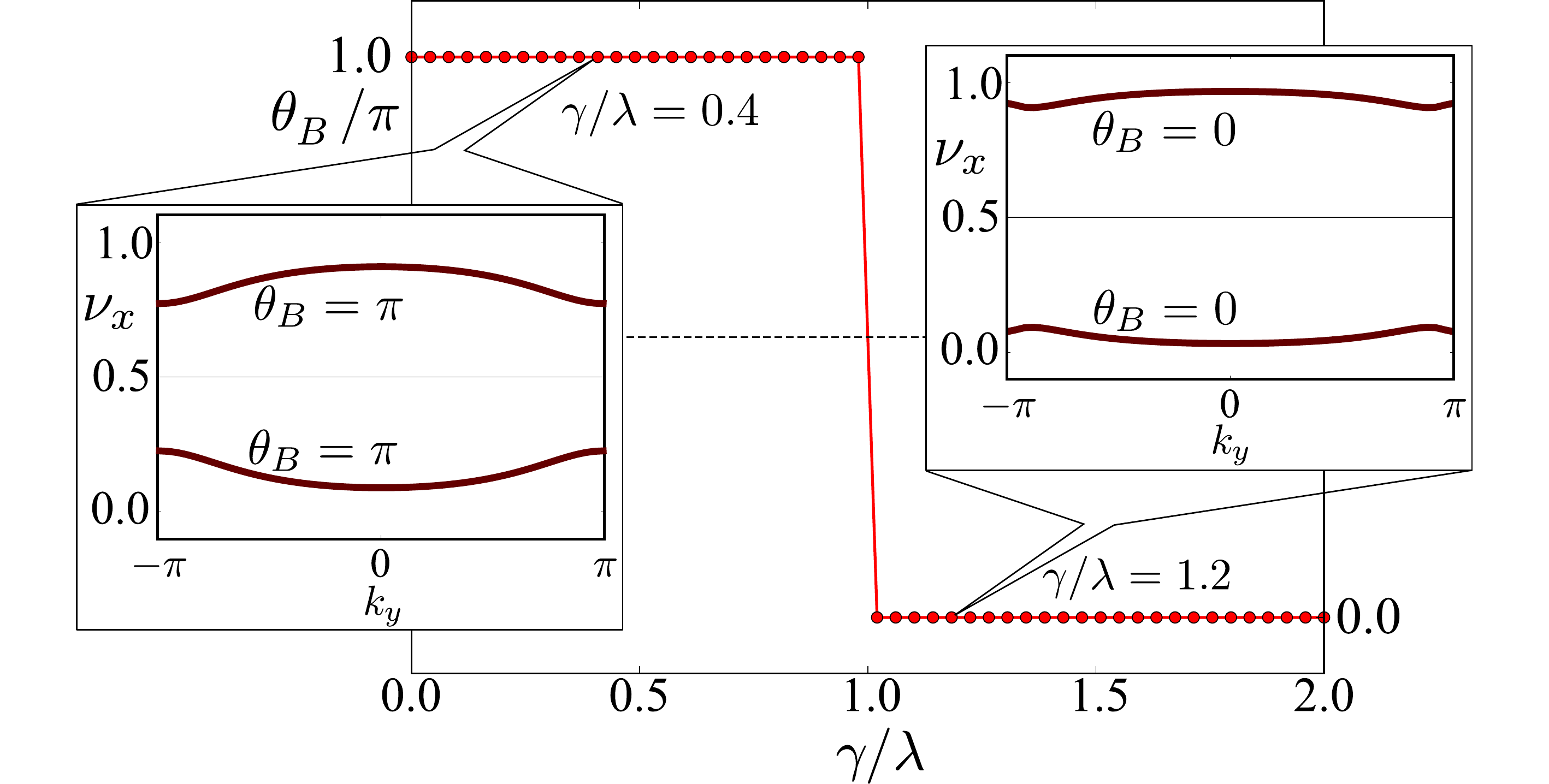}
	\caption{{\bf Bosonic topological invariants.}  
			The Berry phase $\theta_B/\pi$ of the mechanical SOTI model
			calculated from the supersymmetric Berry curvature \eqref{eq:SUSYBerryCurvature} 
			versus the staggering parameters $\gamma/\lambda$.			
			The insets display the eigenvalues $\nu_x$ of the edge Hamiltonian $\mathcal{H}_{x,{\bf k}}$ plotted against $k_y$ in the topological ($\gamma/\lambda<1$) 
			and the trivial ($\gamma/\lambda>1$) phase in which the two bands acquire distinct Berry phases of  
			$\theta_B=\pi$ or $\theta_B=0$, respectively.}
\label{fig:BosonicInvariants}
\end{figure}

Let us apply this reasoning to the SOTI balls and springs model
and demonstrate that its corner modes owe their existence to the topology of the bulk excitation spectrum, shown in Fig.~\ref{fig:SOTI}. 
In order to decode the topology, we consider bosonic Wilson loop operators (see appendix~\ref{app:topologyofcornermodes}), 
which, using the formalism above, we define in  terms of the Berry curvature ${\mathcal A}_{\rm SUSY}$ of Eq.~\eqref{eq:SUSYBerryCurvature}.
The Wilson loop operators~\cite{hughes-bernevig_quadrupole_insulators} $\mathcal{W}_{x,{\bf k}}$ and $\mathcal{W}_{y,{\bf k}}$ along the $k_x$ and $k_y$-directions, respectively,
 define edge Hamiltonians $\mathcal{H}_{x,{\bf k}}$ and $\mathcal{H}_{y,{\bf k}}$ \cite{fidkowski2011model} which can be adiabatically connected to the physical Hamiltonian of the $x$ or $y$ edge of the mechanical model. 
For $\gamma/\lambda<1$, 
each eigenfunction of $\mathcal{H}_{x/y,{\bf k}}$ acquires a Berry phase of $\pi$ as illustrated in Fig.~\ref{fig:BosonicInvariants} -- this is
 the topological phase, which is endowed with the feature of localized corner modes.
For $\gamma/\lambda>1$, each eigenfunction of $\mathcal{H}_{x/y,{\bf k}}$ acquires a Berry phase of zero (i.e. this is a trivial phase), 
thus demonstrating the existence of two topologically distinct phases of the mechanical model, as summarized in Fig.~\ref{fig:BosonicInvariants}.

%%%%%%%%%%%%%%%%%%%%%%%%%%%%%%%%%%%%%%%%%%%%%%%%%%%%%%%%%%%%%%%%%%%%
% Outlook
%%%%%%%%%%%%%%%%%%%%%%%%%%%%%%%%%%%%%%%%%%%%%%%%%%%%%%%%%%%%%%%%%%%%

\noindent {\em Outlook.--}
The fundamental link between topological mechanics and supersymmetry, which we have explicitly formulated in this manuscript, 
provides both conceptual and practical insights. 
On the conceptual side, it is the definition of topological invariants for mechanical (bosonic) systems via
the supersymmetric Berry curvature \eqref{eq:SUSYBerryCurvature}, which allows for a hitherto unexplored perspective
on these systems that might lead to a deeper understanding of the classification of bosonic insulators.
In expanding the conceptual connection in the future, it would be interesting to go beyond the single-particle equivalences
employed in the current work and use SUSY to connect many-fermion states with mechanical (bosonic) analogues.
In practical terms, our SUSY construction allows to translate many of the topological Majorana band theories
to topological balls and springs models ready to be built in the lab. 
As demonstrated for the mechanical Kitaev model and SOTI, the obtained mechanical models might be of intriguing simplicity.
\footnote{
However, some of the complexity of the original models might still be hidden in the occurrence of {\em negative} spring constants, 
which might impede a physical realization in the lab. We note, however, that negative spring constants do not pose an absolute no-go limit, since they can in fact be realized approximately as discussed in Appendix \ref{app:negativesprings}. 
}
One interesting future avenue to explore is whether the mechanical analogues of Majorana fermion systems discussed in this work, 
can regain their quantum mechanical character, e.g. by employing optomechanical systems or nano-scale metamaterials \cite{Aspelmeyer2014}, and thereby
allow to realize the bosonic quantum modes that are SUSY-partners of Majorana fermions.

%%%%%%%%%%%%%%%%%%%%%%%%%%%%%%%%%%%%%%%%%%%%%%%%%%%%%%%%%%%%%%%%%%%%
% Acknowledgments
%%%%%%%%%%%%%%%%%%%%%%%%%%%%%%%%%%%%%%%%%%%%%%%%%%%%%%%%%%%%%%%%%%%%

%\begin{acknowledgments}
\noindent
{\em Acknowledgments.---}
We thank V. Dwivedi, A. Rosch, and in particular M. Zirnbauer for enlightening discussions. 
We gratefully acknowledge the hospitality of the Kavli Institute for Theoretical Physics, supported by NSF PHY-1125915, 
where this work was initiated during the ``Intertwined orders" program.
The Cologne group acknowledges partial funding from the DFG within CRC 1238 (project C02) and CRC/TR 183 (project B01).
The numerical simulations were performed on the CHEOPS cluster at RRZK Cologne employing the julia package DifferentialEquations.jl \cite{DifferentialEquations.jl}.
%\end{acknowledgments}

%%%%%%%%%%%%%%%%%%%%%%%%%%%%%%%%%%%%%%%%%%%%%%%%%%%%%%%%%%%%%%%%%%%%
% Bibliography
%%%%%%%%%%%%%%%%%%%%%%%%%%%%%%%%%%%%%%%%%%%%%%%%%%%%%%%%%%%%%%%%%%%%

\bibliography{SUSYconstruction}

%%%%%%%%%%%%%%%%%%%%%%%%%%%%%%%%%%%%%%%%%%%%%%%%%%%%%%%%%%%%%%%%%%%%
% Appendices
%%%%%%%%%%%%%%%%%%%%%%%%%%%%%%%%%%%%%%%%%%%%%%%%%%%%%%%%%%%%%%%%%%%%

\appendix

%%%%%%%%%%%%%%%%%%%%%%%%%%%%%%%%%%%%%%%%%%%%%%%%%%%%%%%%%%%%%%

\section{Building springs with negative k}
\label{app:negativesprings}

A recurring element in our SUSY construction of mechanical balls and springs models with non-trivial topological properties
are springs with {\em negative} spring constants. The physical realization of such springs might be considered elusive, since an isolated spring with negative spring constant $k$ is inherently unstable -- for any infinitesimal deviation from the rest position, such a spring would exert a force that further pushes away from this rest position, with the strength of the force further increasing with distance. 
Nevertheless, as we will discuss in this Appendix one can in fact construct springs that indeed exhibit negative spring constants, 
at least in a limited range, and thereby approximate the desired behavior. In the following, we identify and discuss two simple settings for such ``approximate" springs with negative spring constant.

%%%%%%%%%%%%%%%%%%%%%%%%%%%%%%%%%%%%%%%%%%%%%%%%%%%%%%%%%%%%%%

\subsection{Using conventional springs}

\begin{figure}
	\centering
	\includegraphics[width=0.95\columnwidth]{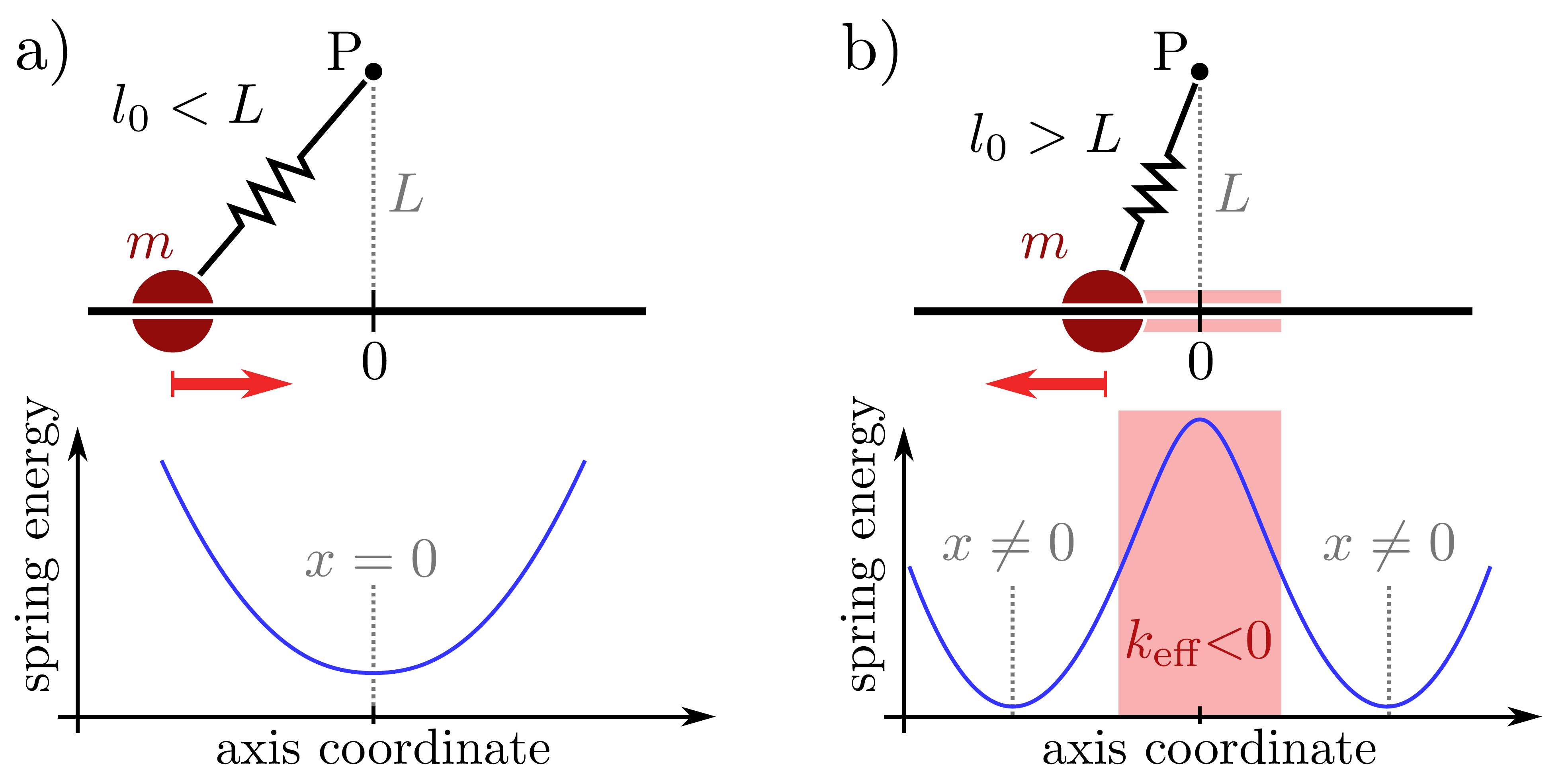}
	\caption{{\bf Negative spring constant from conventional springs.} 
			Sketch of a spring constrained between point P and a free sliding mass $m$ on an axis (at distance $L$ from P). 
			Depending on the rest length $l_0$ of the spring  in relation to the distance $L$
			the effective spring constant in the vicinity of $x=0$ has either positive sign as depicted in a) or negative sign as 				depicted in b).
			The bottom panels show the spring energy as function of the axis coordinate.}
	\label{fig:negative-k-springs}
\end{figure}

First, consider a conventional spring (of constant spring coupling $k$) being attached to both an arbitrary point $P$ and a mass $m$ which is fixed to slide freely on an axis in distance $L$ to $P$, see the illustration in Fig.~\ref{fig:negative-k-springs}. For rest lengths of the spring $l_0 < L$, corresponding to the scenario in Fig.~\ref{fig:negative-k-springs}(a), the mass will be pulled to a rest position on the axis which has the smallest distance to  $P$ which can be denoted by the axis coordinate system as $x=0$. If one increases the rest length of the spring to $l_0 > L$, corresponding to the scenario in Fig.~\ref{fig:negative-k-springs}(b), the rest position of the mass changes as $x=0$ becomes unstable. The stable minimum is now given by some $x \neq 0$. However, since $x=0$ is an unstable rest point, Taylor-evolving its energy around $x=0$ yields $E \sim -x^2 + \mathcal{O}(x^3)$ (since the $\mathcal{O}(x)$ term vanishes because of the rest property and the sign comes the fact that $x=0$ is repulsive). This form of energy functional directly shows that in the vicinity of $x=0$ the system behaves like a spring with negative spring constant.

%%%%%%%%%%%%%%%%%%%%%%%%%%%%%%%%%%%%%%%%%%%%%%%%%%%%%%%%%%%%%%

\subsection{Using gravity}

Second, consider a one-dimensional rope of mass-density $\rho$ resting frictionless inside a U-shaped pipe and a mass $m$ attached to one of its ends. The rope is bend by the U-bend and its right and left end reach up to heights $x_R$ and $x_L$. The potential energy of the system will be given by
\begin{equation}
	E = mgx + \int_0^{x_R} g\rho x' \; dx' + \int_0^{x_L} g\rho x' \; dx' \,.
\end{equation}
Using the additional constraints of total rope length $L = x_R + x_L$, constant mass density $\rho$, and the position $x$ of the mass $m$ being fixed to $x=x_R$, the energy functional becomes
\begin{equation}
	E = mgx + \frac{1}{2} g\rho \left(x^2 + (L-x)^2\right) \,.
\end{equation}
This energy functional is quadratic in $x$ with a positive curvature, i.e. the system behaves just like a regular spring around the rest position  for which the forces in both legs of the U-pipe balance.
However, the sign of the spring constant depends ultimately on the sign of $g\rho$ and can therefore be flipped by flipping the U-pipe (just as if the direction of $g$ relative to the direction of $x$ is reversed).
In such a device, gravity can therefore be used to build a spring-like system with any sign and strength of its spring constant by adjusting mass density $\rho$ of the rope and orientation of the enclosing U-pipe.

\begin{figure}
	\centering
	\includegraphics[width=0.95\columnwidth]{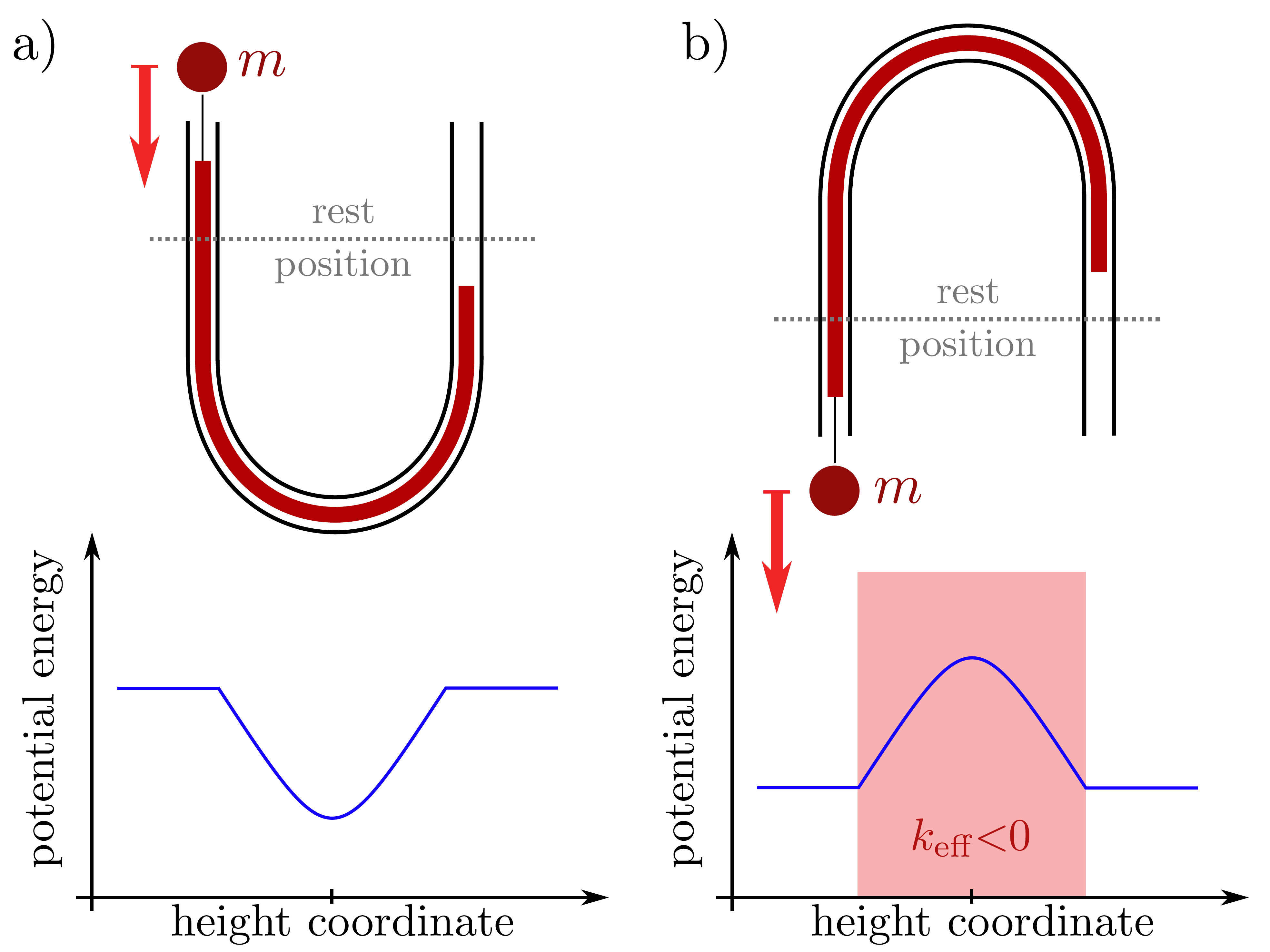}
	\caption{{\bf Negative spring constant from gravity.}
		     Sketch of  massive rope resting frictionless inside a (a)  U-shaped pipe and (b) a flipped U-shape pipe.
		     The lower panels show the potential energyas function of mass coordinate, indicating an $E \sim \pm x^2$ form 
		     in the vicinity of $x=0$ -- reminiscent of a spring with positive/negative spring constant, respectively.
		     }
	\label{fig:negative-k-gravity}
\end{figure}

%%%%%%%%%%%%%%%%%%%%%%%%%%%%%%%%%%%%%%%%%%%%%%%%%%%%%%%%%%%%%%

\section{Corner mode simulations}
\label{app:cornermodes}

The topologically non-trivial feature of the mechanical SOTI system, discussed in the main text, is the occurrence of `floppy' corner modes, i.e. zero-frequency modes located at the corner sites whose existence is protected by the bulk topology of the SOTI.
Since the mechanical model is defined for only one of the two sublattices (denoted B in the main text) and further decomposes into two decoupled parts (denoted B1 and B2 in the main text), every one of theses four subparts effectively contains only one of the four corner modes, located at the corner site that can be identified one-to-one with a corner site of the original fermion model (whereas the other three corner sites correspond to bulk sites in the original fermion model).

\begin{figure}[t]
	\centering
	\includegraphics[width=\linewidth]{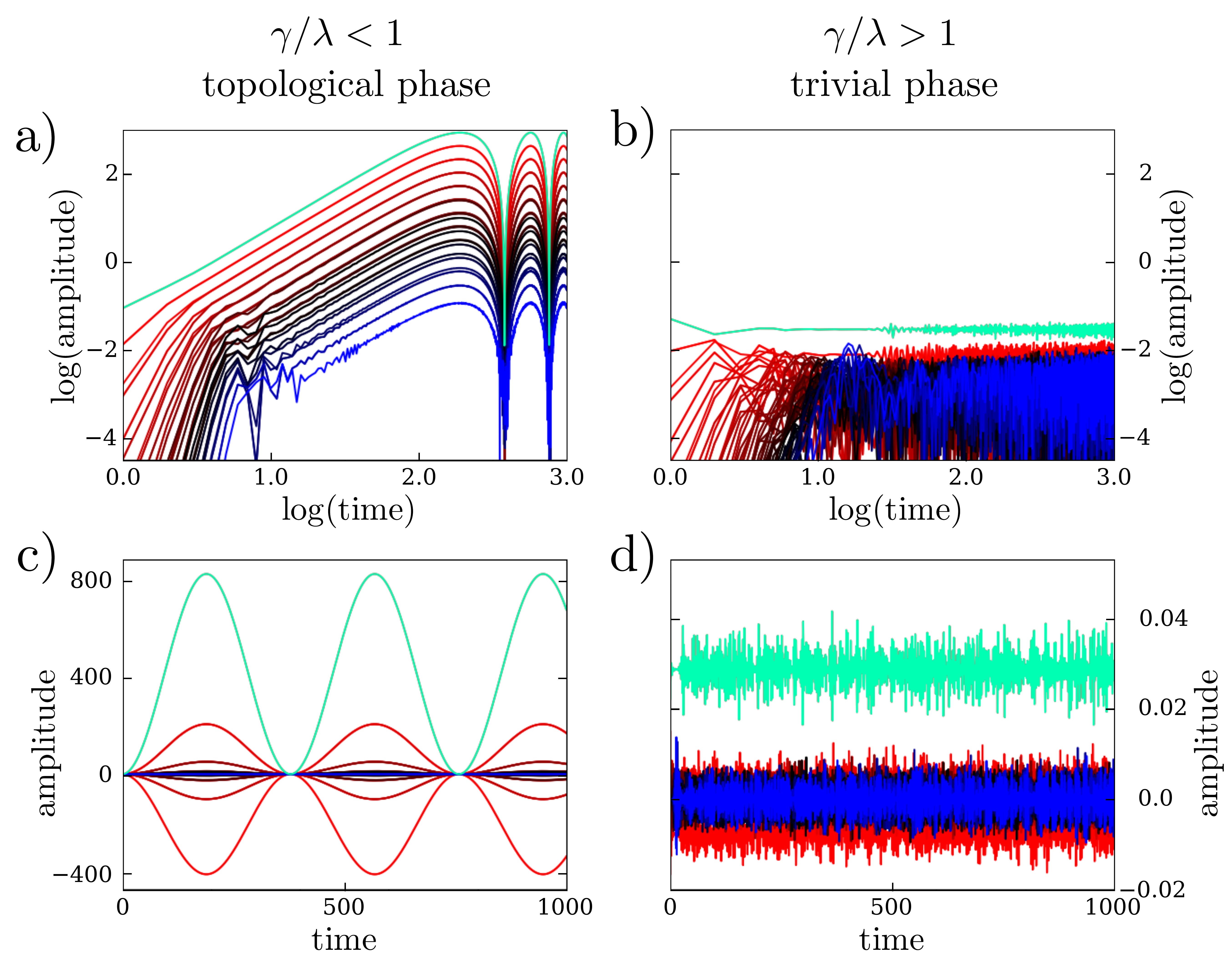}
	\caption{{\bf Corner modes} in a balls \& springs SOTI model a),b) Comparison between logarithmic coordinate amplitudes obtained in numerics in the topological phase a) and the trivial phase b) with small force exerted on the corner. A corner mode is visible in the topological phase a) as a collective free fall motion in time which is exponentially localized in the corner (green line: driven site, red to blue: increasing distance to corner-mode center). The mode has a small but finite frequency. Lower panels c) and d) show the same data on non-logarithmic axes. Low-frequency oscillations in the topological phase c) are clearly distinct from the absence of response in the trivial phase.}
	\label{fig:corner}
\end{figure}

To probe the floppy corner mode numerically, we switch from the periodic finite-frequency drive (used to calculate the bulk energy spectra) to a ``static zero-frequency drive", i.e. we exert a force which is constant in time on only the corner site. For a corner with a floppy corner mode, there is no restoring force and the corner mass is subject to a constant acceleration and thus performs a free-fall like motion in time. The resulting amplitudes under static drive can be seen in Fig.~\ref{fig:corner} where we demonstrate the occurrence of the corner-mass response in the topological phase as well as the absence of any specific corner response in the trivial phase for a system of $9 \times 9$ masses.
Our numerical results show that for a wide range of force amplitudes as well as different driven sites, the same corner motion can be excited, indicating a collective mode.
For any finite quantum system the SOTI corner modes are expected to be of finite energy which translates to a finite frequency in our mechanical model.
We can confirm this finite frequency behavior and its scaling by our numerics as we observe periodic behavior on large timescales which scale by the system size $L$ as $\tau \sim 2^{L}$.

%%%%%%%%%%%%%%%%%%%%%%%%%%%%%%%%%%%%%%%%%%%%%%%%%%%%%%%%%%%%%%
% topology of mechanical SOTI 
%%%%%%%%%%%%%%%%%%%%%%%%%%%%%%%%%%%%%%%%%%%%%%%%%%%%%%%%%%%%%%

\section{The topology of mechanical SOTI}
\label{app:topologyofcornermodes}

In this Appendix, we present an explicit computation of the topological invariant that characterizes the topological phase of the mechanical SOTI and demands the existence of `floppy' corner modes. The corresponding SUSY partner represents a fermionic SOTI consisting of Majorana fermions as noted in the main text. The mechanical model can be specified via its rigidity matrix of the form
\begin{equation}
  {\bf R}
    =
    \begin{pmatrix}
      \mathbf{1} & \mathbf{0} \\
      \mathbf{0} & \mathbf{A}
    \end{pmatrix},
    \label{Rsoti}
\end{equation}
where the block $\mathbf{A}$  is given, in Fourier space, by 
\begin{equation}
 {\bf A} = 
  \begin{pmatrix}
    \gamma + \lambda e^{-ik_x} & -\gamma - \lambda e^{ik_y} \\
    \gamma + \lambda e^{-ik_y} &  \gamma + \lambda e^{ik_x} \\
  \end{pmatrix}.
 \label{Asoti}
\end{equation}
The corresponding fermionic model allows for hopping of Majorana fermions on a square lattice in presence of $\pi$-flux per square plaquette with the interplaquette and intraplaquette hopping strengths are denoted $\lambda$ and $\gamma$, respectively. 

The Hamiltonian of the mechanical model in  Fourier space is given by
\begin{equation}
 H_{\rm mech} = 
 \begin{pmatrix}
  p_i & q_i
 \end{pmatrix}
 {\bf R}^\dagger{\bf R}
 \begin{pmatrix}
  p_i \\ 
  q_i \\
 \end{pmatrix} \,,
 \label{Hambos}
\end{equation}
with the form of ${\bf R}$ given in Eqs.~\eqref{Rsoti} and \eqref{Asoti}. Using a compact notation $x_{i\mu}\equiv(p_i~~q_i)^T$, the equations of motion can be expressed as $\dot{x}_{i\mu}=[x_{i\mu},H_{\rm mech}]$. Exploiting the anticommutation relations of $p_i$ and $q_i$ (which are canonically conjugate), we find
\begin{equation}
 \dot{x}_{i\mu}= (\sigma_2 {\bf R}^\dagger{\bf R})_{i\mu,j\nu}~{x}_{j\nu} \,.
 \label{eqom}
\end{equation}
Note the appearance of Pauli matrix $\sigma_2$ from the commutation relation 
\[
	[x_{i\mu},x_{j\nu}]=[\sigma_2]_{\mu\nu}\delta_{ij} \,.
\] 
The solutions to the above equation satisfy a symplectic relation 
\begin{equation}
	V^\dagger \sigma_2 V=\sigma_3 \,,
\end{equation}
 while constituting the columns of $V$ (in a deceasing order of the eigenvalues). 
 
 In  Fourier space, ${\bf R}$ is a $4\times4$ matrix, and since the matrix $\sigma_2 {\bf R}^\dagger{\bf R}$ is isospectral with ${\bf R} \sigma_2 {\bf R}^\dagger$ with the latter having a charge-conjugation symmetry, the frequencies, i.e. the eigenvalues of $\sigma_2 {\bf R}^\dagger{\bf R}$, come in pairs \{$\pm \omega$\}. In the specific model where ${\bf R} \sigma_2 {\bf R}^\dagger$ represents a Majorana Hamiltonian $H_{\rm Majorana}$ which describes the $\pi$-flux model mentioned earlier, the frequencies appear doubly degenerate. The (two) physical states are regarded as those with positive frequencies ($\omega>0$). 

The bosonic Hamiltonian in Eq.~\eqref{Hambos}, which is the SUSY partner to $H_{\rm Majorana}$, features corner modes for $\gamma/\lambda<1$ arising as a consequence of the topology of the bulk excitation spectrum \{$\omega$\}. As we show here, this topology can be explored by constructing Wilson loop operators in terms of the physical eigenstates. 

For any bosonic model specified by a rigidity matrix ${\bf R}$, the Wilson loop $\mathcal{W}_{x,{\bf k}}$ along $k_x$-direction starting from a base point ${\bf k}=(k_x,k_y)$ takes the form 
\begin{equation}
 \mathcal{W}_{x,{\bf k}} = G_{x,{{\bf k}+(N_x-1)\Delta{\bf k}}}\cdots G_{x,{\bf k}+\Delta{\bf k}} G_{x,{\bf k}},
\end{equation}
with $G_{x,{\bf k}}$ being a Wilson line element at ${\bf k}$ defined as
\begin{equation}
 [G_{x,{\bf k}}]_{ab} = f_{\bf k} \langle v_a ({\bf k}+\Delta{\bf k})| {\bf R}^\dagger({\bf k}+\Delta{\bf k}) {\bf R}({\bf k}) |v_b({\bf k})\rangle,
 \label{Wilsonline}
\end{equation}
where $f_{\bf k}=[\omega_a({\bf k}+\Delta{\bf k})\omega_b({\bf k})]^{-1/2}$ and $\Delta{\bf k}=(2\pi/N_x,0)$ for $N_x$ lattice sites along $x$. As mentioned before, for the  mechanical SOTI, the eigenstates $v_{a,b}$ considered in Eq.~\eqref{Wilsonline} are doubly degenerate ($a,b=1,2$) with frequency $\omega_a=\omega_b (>0)$. Similarly follows the construction for $\mathcal{W}_{y,{\bf k}}$ along $k_y$. Also note that in the thermodynamic limit ($\Delta{\bf k}\rightarrow 0$), $[G_{x,{\bf k}}]_{ab}=\delta_{ab}$ (similar to the unitary nature of fermionic eigenfunctions) obtained using Eq.~\eqref{eqom} . However, for finite $N_x$, we resort to a singular value decomposition of  $G=U\Sigma V^\dagger$, where $\Sigma$ is a diagonal matrix containing the singular values and proceed with the matrix $UV^\dagger$ to construct unitary Wilson loops $\mathcal{W}_{x/y,{\bf k}}$.

The Wilson loop operator $\mathcal{W}_{x,{\bf k}}$($\mathcal{W}_{y,{\bf k}}$) then defines an edge Hamiltonian $\mathcal{H}_{x,{\bf k}}$($\mathcal{H}_{y,{\bf k}}$) via
\begin{equation}
 \mathcal{H}_{x/y,{\bf k}} = (i/2\pi)\log\mathcal{W}_{x/y,{\bf k}} \,,
\end{equation}
which can be adiabatically connected to the physical Hamiltonian of the $x(y)$ edge of the mechanical model and for $\gamma/\lambda<1$($>1$) represents a  topological (trivial) fermionic insulator. The eigenvalues of $\mathcal{H}_{x,{\bf k}}$($\mathcal{H}_{y,{\bf k}}$) [denoted as \{$\nu_x$\} (\{$\nu_y$\})] are  functions of $k_y(k_x)$ only. Denoting the eigenstates of $\mathcal{H}_{x,{\bf k}}$, for instance, as $|w_a\rangle$, we observe that in the topological phase  with corner modes, each eigenstate acquires a {\em fermionic Berry phase}
\begin{equation}
  \theta_B=\oint dk_y~\langle w_a({\bf k}) |\partial_{k_y} w_a({\bf k})\rangle = \left\{
  \begin{array}{@{}ll@{}}
    \pi~\forall k_x, & \text{for}\ \gamma/\lambda<1 \\
    0, & \text{otherwise,}
  \end{array}\right.
\end{equation}
defined modulo $2\pi$. 
A similar property holds for $\mathcal{H}_{y,{\bf k}}$ as well, which in combination signify the topological nature of the corner modes observed in the numerical simulations.

\end{document}